*Tuning the flat bands of the kagome metal CoSn with Fe, In, or Ni doping*

*B. C. Sales[1], W. R. Meier[1], A. F. May[1], J. Xing[1], J. -Q Yan[1], S. Gao[1,2], Y. H. Liu[2], M. B. Stone[2], A. D. Christianson[1], Q. Zhang[2],  and M. A. McGuire[1]*

[1]*Materials Science and Technology Division, Oak Ridge National Laboratory, Oak Ridge TN 37831*
[2]*Neutron Scattering Division, Oak Ridge National Laboratory, Oak Ridge TN 37831*

Abstract

CoSn is a Pauli paramagnet with relatively flat d-bands centered about 100 meV below the Fermi energy, $E_F$.  Single crystals of CoSn lightly doped with Fe, In or Ni are investigated using x-ray and neutron scattering, magnetic susceptibility and magnetization, ac susceptibility, specific heat and resistivity measurements. Within the rigid band approximation, hole doping with a few percent of Fe or In should move the flat bands closer to $E_F$, whereas electron doping with Ni should move the flat bands further away from $E_F$. We provide evidence that this indeed occurs. Fe and In doping drive CoSn toward magnetism, while Ni doping suppresses CoSn's already weak magnetic response. The resulting ground state is different for Fe versus In doping. For Fe-doped crystals, $Co_{1-x}Fe_xSn$, with 0.02 < x <0.27, the magnetic and specific heat data are consistent with the formation of a spin glass, with a glass temperature, $T_g$, ranging from 1 K for x =0.02 to 10 K for x = 0.27. Powder and single crystal neutron diffraction found no evidence of long-range magnetic order below $T_g$ for samples with x ≈ 0.17. For In-doped crystals, $CoSn_{1-y}In_y$, both the magnetic susceptibility and the Sommerfeld coefficient, $\gamma$, increase substantially relative to pure CoSn, but with no clear indication of a magnetic transition for 0.05 < y < 0.2. CoSn crystals doped with Ni ($Co_{0.93}Ni_{0.07}Sn$) have a significantly smaller magnetic susceptibility and $\gamma$ than pure CoSn, consistent with flat bands further from $E_F$.

Introduction

Using the geometry of crystal lattices to generate narrow relatively flat electronic bands in reciprocal space is an interesting new approach to generating strong correlations and potentially novel emergent phenomena.  The most spectacular example of this approach is the moire lattice produced in twisted graphene bilayers. Upon doping using a gate voltage, this lattice of carbon atoms generates a phase diagram reminiscent of the cuprate superconductors [1]. This suggests that the geometry of the crystal lattice may be capable of generating electron correlations without elements that contain partially filled d or f shells, which is very appealing. Theoretical calculations suggest that similar effects may occur in Lieb, dice and kagome lattices [2,3, see Fig. 1c]. The basic idea is that the destructive interference caused by the different electron hopping paths on these lattices results in electron localization and very narrow bands that have little dispersion in k-space- i.e. flat bands. Real crystals are 3D, but there are quasi 2D layered compounds where each layer has the desired geometrical arrangement of atoms. The crystal structures that have kagome layers include $Fe_3Sn_2$ [4,5], $Co_3Sn_2S_2$ [6], CoSn family [7-13], the $HfFe_6Ge_6$ family [14] and the recently reported $AV_3Sb_5$, with A = K, Rb or Cs. [15-17]. For the flat bands to have a maximum impact on properties, the bands should be tuned close to the Fermi energy, $E_F$. Tight-binding calculations that only consider nearest neighbor interactions produce perfectly flat bands as illustrated in Fig. 1c. However, more complete density functional theory (DTF) calculations that include spin-orbit coupling modify this simple picture and result in d bands with a finite spread in energy, as illustrated in Fig. 1d. From DFT calculations for CoSn [13], the estimated width of the flat d-bands nearest $E_F$ is about 0.3 eV. The width of the flat d-bands results in a broadened peak in the density of states centered below $E_F$ (Fig. 1d) but with some contribution from the tail of the peak extending up to $E_F$.

The structure of CoSn (*P6/mmm*, 191) consists of kagome nets of Co with the hexagonal holes filled with Sn as shown in Fig. 1a and Fig. 1b. The kagome layers are stacked along the *c* axis separated by layers of Sn. Relatively flat electronic bands are centered about 100 meV below the Fermi energy as determined by angle resolved photoelectron spectroscopy (ARPES)[11,12], density functional theory (DFT) [11,12,13, also see Fig. 1d] and magnetic susceptibility data [13]. CoSn is a Pauli paramagnet and no phase transition is observed in CoSn down to 0.4 K and no superconductivity is found down to 0.1 K. The isostructural compound FeSn is an antiferromagnet with a Neel temperature, $T_N$, of 365 K and $Co_{1-x}Fe_xSn$ was found to maintain long range antiferromagnet order for x>0.4, although the antiferromagnetic order switches between planar and axial [9]. This behavior is illustrated in Fig. 2, where data from small x from the current study are also included to complete the phase diagram.

In the present work we explore the effects of low concentrations of Fe, In and Ni on the position of the d bands in CoSn with respect to the Fermi energy using x-ray and neutron scattering, magnetic susceptibility, magnetization, ac susceptibility, specific heat and resistivity measurements. Hole doping with a few percent of Fe or In should move the flat bands closer to $E_F$, whereas electron doping with Ni should move the Fermi energy further away from the flat bands. We find that $Co_{1-x}Fe_xSn$ crystals with $x \approx 0.02$ results in a spin glass phase with $T_g \approx 1$ K, and $T_g$ increases linearly with Fe concentration up to about x = 0.3 where $T_g \approx 10$ K. In addition to adding holes to CoSn, each Fe atom also adds a relatively local magnetic moment with S=1. The combination of a low concentration of localized Fe spins, plus an enhanced density of states results in a magnetic feature with the characteristics of a spin glass. This spin glass phase persists up to values of $x \approx 0.55$, which is close to the percolation limit of $p_c = 0.524$ for a kagome lattice [18]. For compositions $0.4 < x < 0.6$, there are two magnetic transitions, with long range antiferromagnetic order present from approximately $0.4 < x < 1.0$. The spin glass phase likely results from a different type of magnetic exchange coupling than occurs for the more concentrated Fe alloys of $Co_{1-x}Fe_xSn$ with x > 0.4. A possible quantum critical point for values of x near 0.4 is avoided by the emergence of this spin-glass phase that persists nearly to pure CoSn, at least to x ≈ 0.02.

By contrast, hole doping with In, $CoSn_{1-y}In_y$, produced no magnetic order above 0.4 K at the maximum solubility for In of y ≈ 0.2. While In doping substantially increases the density of states, it is apparently not enough to drive magnetic order or superconductivity. Hole doping with In, however, does quadruple the density of states at $E_F$ as determined from low temperature specific heat measurements and increases the magnetic susceptibility by an order of magnitude, consistent with the picture of tuning the flat bands closer to $E_F$. CoSn crystals doped with Ni ($Co_{0.93}Ni_{0.07}Sn$) have a significantly smaller magnetic susceptibility and γ than pure CoSn, consistent with flat bands further from $E_F$.

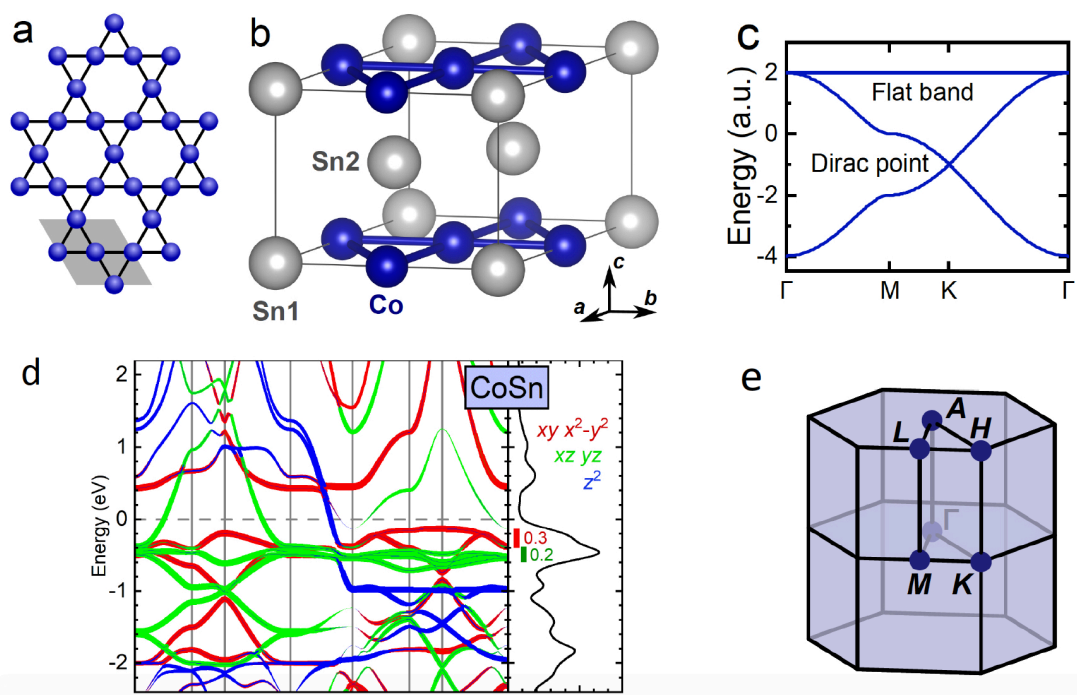

Fig.1 (a,b) Crystal structure of CoSn emphasizing the kagome planes of Co atoms (c) electronic structure of a kagome layer calculated using the tight binding model, which assumes only nearest neighbor interactions (d) electronic structure of CoSn as calculated using DFT with spin orbit coupling, reproduced from our previous work (ref 13). Bands with thicker lines have higher 3d orbital character, and the type of 3d orbital is indicated by the color. The relatively flat 3d bands with dominant xy or $x^2$-$y^2$ character (highlighted in red) are closest to the Fermi energy. The density of states is shown on the right edge of the panel (e) Standard notation for the high symmetry points in the CoSn Brillouin zone.

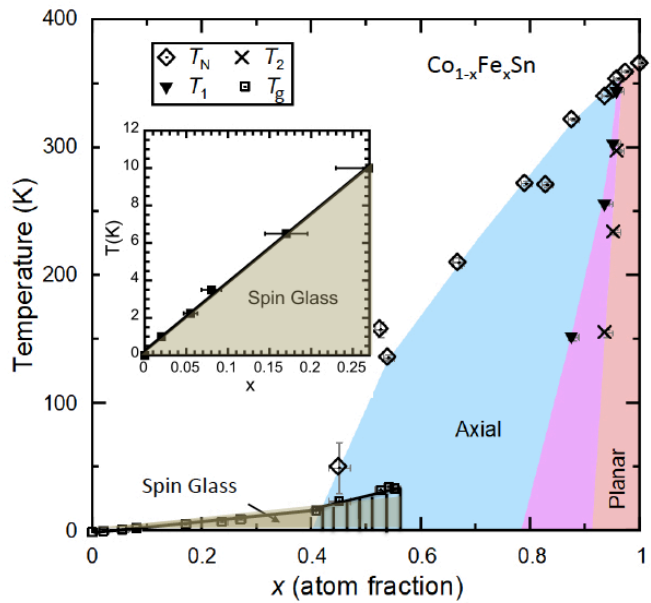

Fig 2. Combined phase diagram determined from specific heat, magnetic susceptibility, ac susceptibility, and neutron scattering data for single crystals of $Co_{1-x}Fe_xSn$ with $0<x<1$ including the data from [9] and the present work highlighted in the inset. The regions labeled "planar" or "axial" correspond to A-type antiferromagnetic order with the Fe spins aligned in the kagome planes or along the *c* axis, respectively. In the purple area of the phase diagram, the spins are in a "tilted" configuration. The spin glass transition is labeled $T_g$ in Fig. 1 and is defined as the temperature where the zero-field cooled and field-cooled susceptibility data diverge in an applied field of 100 Oe. A possible quantum critical point near x=0.4 is avoided by the emergence of the spin glass phase. For compositions $0.4 < x < 0.6$, there are two magnetic transitions. The spin glass phase disappears near the percolation threshold for a kagome lattice of $p_c$ = x = 0.524, [18] although long-range antiferromagnetic order persists to lower values of x ≈ 0.4.

**Crystal growth and chemical characterization**

Crystal growth of Fe doped CoSn alloys from a Sn flux have been described in detail previously and a Ni doped crystal, $Co_{0.93}Ni_{0.07}Sn$ was grown with a similar method. [8,9,13] Indium has a limited solubility in $CoSn_{1-y}In_y$ of about y ≈ 0.2. Crystals with a composition of $CoSn_{0.95}In_{0.05}$ were grown from a starting composition of 20g Sn, 10g In, and 1.2 g Co in a 10cc alumina crucible sealed inside an evacuated silica ampoule. After homogenizing the liquid at 1130 °C for 24 h, and physically shaking the liquid at 900 °C, the melt was cooled at 1°C/h to 620 °C. The silica tube was removed from the furnace and the excess In-Sn flux was centrifuged into a 10 cc catch crucible filled with quartz wool. The crystals grew as hexagonal bars a few mm on a side and up to 20 mm in length. The chemical composition of the crystals was determined using a Hitachi TM-3000 scanning electron microscope and a Bruker Quantax 70 energy dispersive X-ray (EDX) attachment. For the larger Fe-doped crystals careful EDX measurements indicated the Fe concentration of some of the crystals was typically about 10-15% higher at the center the hexagonal face than at the edge, but that the chemical composition along the length of the crystals was relatively uniform. The average chemical composition is used to describe the composition of each crystal. Although not the focus of the present work, there was concern that the two magnetic transitions observed for values of 0.4 < x < 0.6 (see Fig. 2) could be due to different Fe concentrations in different parts of the crystal. To check this possibility we selected a small 1 mg crystal with average composition $Co_{0.41}Fe_{0.59}Sn$ that showed both magnetic transitions ($T_N \approx 163$ K, Tg $\approx 40$ K). EDX measurements both radially and along the length of this crystal showed less than a 2% variation in the measured chemical composition. Based on this and similar data from other crystals, we believe the two magnetic transitions for 0.4 < x < 0.6 are not due to chemical inhomogeneity. Magnetic susceptibility, ac susceptibility, resistivity and specific heat measurements were made using commercial equipment from Quantum Design. For the Fe-doped CoSn crystals with Fe concentrations less than x = 0.2, the magnetic susceptibility data between 15 K and 300 K (Fig. 3a) are accurately fit by a Curie-Weiss law plus a small constant susceptibility ($C/(T+\Theta) +\chi_0$) . If it is assumed that only Fe contributes to the CW portion of the susceptibility, the Fe concentration can be extracted from the value of C. Remarkably good agreement between the average Fe concentration measured with EDX and the value determined from C was obtained if it was assumed that each Fe had S=1 and g=2. This is illustrated in Fig. 3 with additional information in the caption. A nice feature of the correlation is the ability to use C to accurately determine small Fe concentrations that are difficult to measure using EDX.

The lattice constants and phase purity were determined from powder x-ray diffraction using a PANalytical X'pert Pro diffractometer with a Cu K$\alpha$ tube with an incident beam monochrometer. The lattice constants versus composition are given in Fig 4. In general, the variation of the lattice constants with Fe content are in agreement with previous reports [9, 19], although at low Fe doping the *a* lattice constant may actually decrease slightly relative to pure CoSn before increasing with larger Fe concentrations. The addition of In, increases both *a* and *c* lattice constants by a small amount while the addition of Ni increases *c* slightly but decreases *a*.

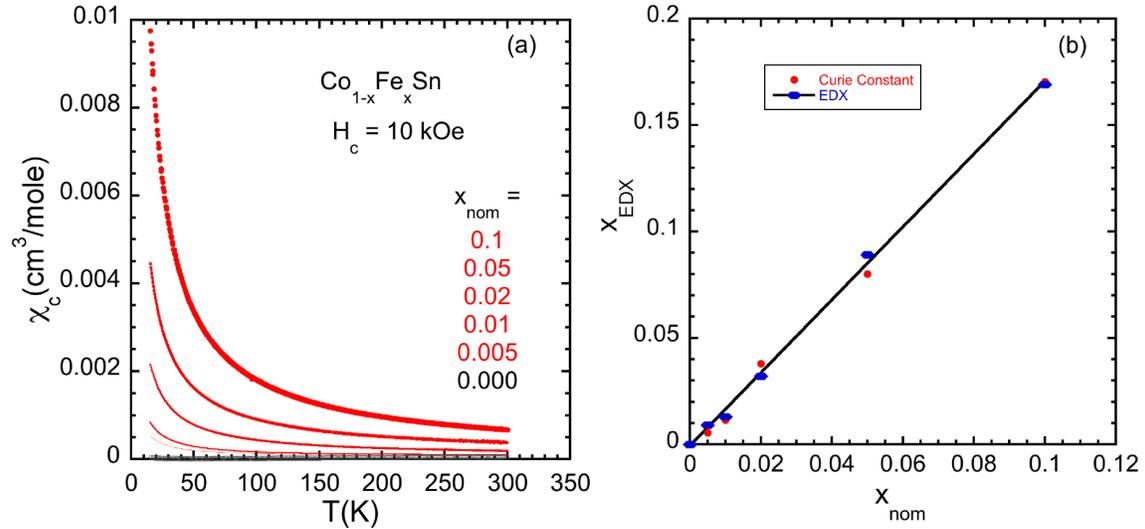

Fig. 3 (a) Magnetic susceptibility versus temperature for a series of Fe-doped CoSn crystals from 15 to 300 K, where $x_{nom}$ refers to the amount of Fe added relative to Co at the start of the crystal growth process. (b) If the data in (a) are fit to a Curie-Weiss law, $C/(T+\theta) + \chi_0$, the Curie constant, C, should be proportional to the Fe concentration if only the Fe atoms are magnetic. Assuming S=1 for each iron, this gives an effective moment of 2.82 $\mu_B$ for each Fe atom and the red points in Fig 3b. The blue points are the actual Fe content as determined from careful EDX measurements on the same crystals used to produce the data shown in Fig. 3a. The agreement between the two sets of data are within experimental error. The fitted values for C, $\theta$, and $\chi_0$ are given in Table 1.

Table 1. Fit of the susceptibility data in Fig 3a to $C/(T + \theta) + \chi_0$, and comparison to the Fe concentration, $x_{CW}$ , extracted from the value of C with the value, $x_{EDX}$ determined from EDX measurements on the same crystals.

| $Co_{1-x}Fe_xSn$ $x_{nom}$ | $x_{CW} =$ $8C/(2.82)^2$ | $\theta(K)$ | $\chi_0$ $cm^3/mole$ | $x_{EDX}$ |
|---|---|---|---|---|
| 0 | - | | | 0.00 |
| 0.005 | 0.0056 | -3.7 | 8.27e-5 | 0.009 |
| 0.01 | 0.012 | -0.66 | 6.32e-5 | 0.013 |
| 0.02 | 0.038 | 2.94 | 7.16e-5 | 0.032 |
| 0.05 | 0.08 | 3.35 | 12.8e-5 | 0.089 |
| 0.1 | 0.170 | 2.41 | 12.6e-5 | 0.17 |

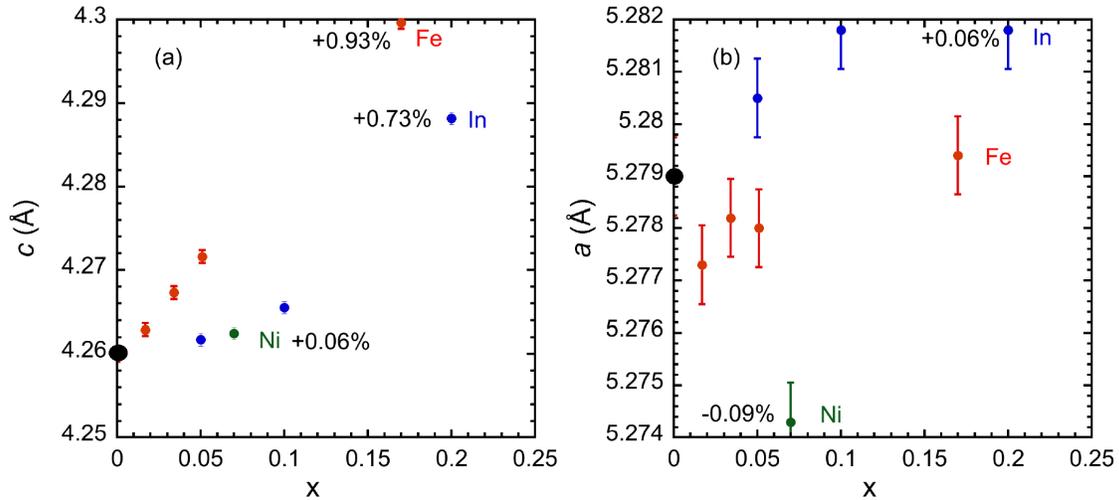

Fig. 4. Room temperature lattice constants for Fe, In and Ni doped crystals, as determined from X-ray powder diffraction. The value x is the EDX measured atomic fraction of each dopant i.e. $Co_{1-x}Fe_xSn$, $Co_{1-x}Ni_xSn$, or $CoSn_{1-x}In_x$. The lattice constants for pure CoSn are shown by the filled black circle.

## Results

### *Fe-doped CoSn crystals*

Magnetic susceptibility data for several CoSn crystals doped with concentrations of Fe ranging from x = 0.008 to 0.17 are shown in Fig 5a. A broad maximum in the susceptibility coupled with a separation between the zero-field-cooled (zfc) and field cooled (fc) data suggest a type of magnetic transition with the transition temperatures ranging from ≈1 to 6K. A slightly larger susceptibility with the magnetic field aligned along the *c* axis may indicate that the Fe spins tend to lie in the ab plane (Fig. 5b) as is observed in FeSn [8].

Low temperature magnetization data below $T_g$ with H//c are shown in Fig. 5c for $Co_{1-x}Fe_xSn$ crystals with x = 0.02, 0.08 and 0.17. The M(H) curves show some curvature, but approach saturation much more slowly than would be expected for non-interacting spins. For comparison a Brillouin function for S = 1 and g=2 is shown in the figure. Thus, while the linear behavior typical of simple antiferromagnets is not seen, the data does indicate the presence of antiferromagnetic interactions.

Specific heat data from the Fe-doped CoSn crystals (Fig. 6) indicate a low temperature peak consistent with the magnetic susceptibility data as shown in Fig 5a. Using pure CoSn as a reference, the contribution of Fe doping to the specific heat can be extracted. The magnetic susceptibility data shown in Fig. 3a and previous neutron data on pure FeSn indicate S=1 and g=2 per Fe in this family of alloys [8]. The entropy associated with a concentration x of spins with S=1 is xRln(3), where R is the gas constant. Integrating the C/T data in Fig. 6a indicates that only about 60% of this entropy is removed due to the magnetic transition, suggesting a degenerate ground state. Curiously, the amount of entropy removed is very close to xRln(2).

The effect of an external magnetic field on the specific heat of several Fe-doped CoSn crystals was investigated. The typical behavior is illustrated in Fig 6c for a $Co_{0.98}Fe_{0.02}Sn$ crystal. The peak near the transition is suppressed, while the specific heat at higher temperatures is increased. For a magnetic field of H = 80 kOe the increase in C/T near 0.5 K may be due to a nuclear contribution to the heat capacity. Within experimental error there was no effect of a magnetic field on the heat capacity of CoSn.

The broadness of the magnetic transitions coupled with apparent local moment behavior, chemical disorder, separation of zfc and fc data, and significant missing entropy, suggest a degenerate ground state and possible spin glass behavior. Time dependent magnetization and ac susceptibility data shown in Fig. 7, indicate "glassy" behavior. The peak in the real part of the ac susceptibility shifts from 7.85 K to 8.1 K for frequencies of 100 hz and 5000 hz respectively. The magnitude of the imaginary part of the ac susceptibility is much smaller than the real part and increases near $T_g$ and also increases with measuring frequency as expected for a good metal. For the time dependent data shown in Fig. 7c, the sample was cooled in a field of 10 kOe to a temperature below the glass transition temperature and the magnetic field was set to zero, which took about 3.5 minutes, and then the magnetization was measured as a function of time.

Possible long-range magnetic order of compositions near $Co_{0.83}Fe_{0.17}Sn$ was investigated using both single crystal and powder neutron diffraction. Single crystal neutron diffraction experiments were performed on the white beam Laue diffractometer CORELLI at the Spallation Neutron Source (SNS) at $T$ = 1.6 and 15 K in an orange cryostat. A 35 mg single crystal with a composition from EDX of $Co_{0.83}Fe_{0.17}Sn$ was used for the measurements. To obtain a large coverage of the reciprocal space, the crystal was rotated in the horizontal plane over 360° in 1° (2°) steps at 1.6 (15) K, with a counting time of 1 min at each step. 1914 nuclear reflections were collected and reduced to 157 independent reflections for refinement using the FullProf software [20]. The refinement result using the $P6/mmm$ space group gave goodness-of-fit factors of $R_{F2}$ = 7.5 % and $R_{F2w}$ = 6.5 %. The refined composition for Fe is 0.161(1), which is close to the average value determined from EDX measurements. No additional reflection was observed at 1.6 K, indicating the absence of long-range magnetic order. Neutron diffraction experiments on ≈ 4.8 g of powdered crystals of average composition $Co_{0.83}Fe_{0.17}Sn$ were performed on the time-of-fight diffractometer POWGEN at the SNS. Data were acquired at $T$ between 300 K and 1.8 K using an orange cryostat. The 2.667 and 1.5 Å instrumental configurations were employed. No evidence of long-range magnetic order was found, consistent with the single crystal results, and consistent with the hypothesis of a spin glass ground state.

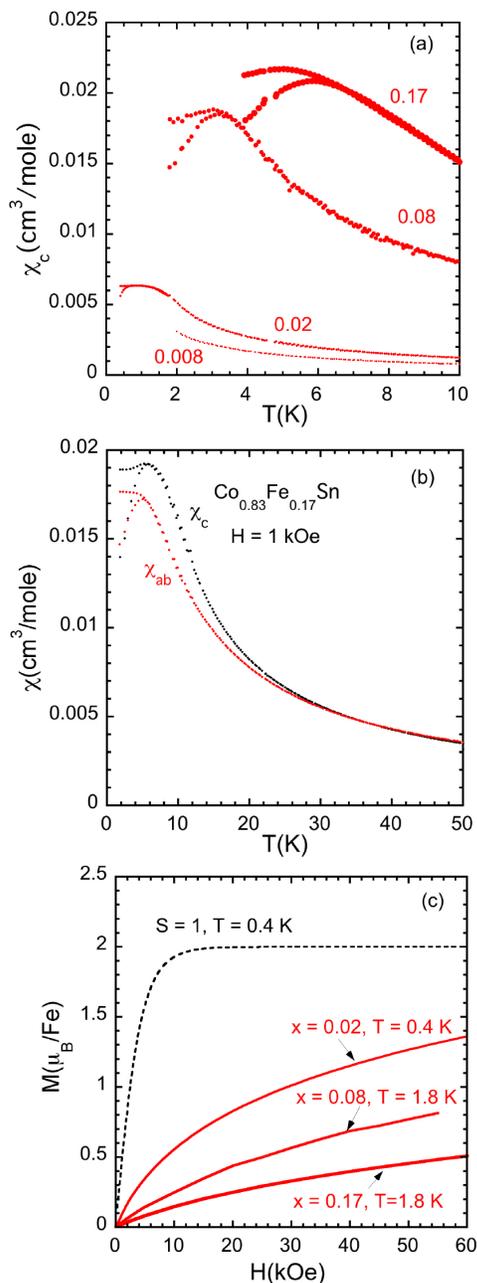

Fig. 5 (a) Zero-field cooled (zfc) and field cooled (fc) magnetic susceptibility data for crystals of $Co_{1-x}Fe_xSn$ with H =100 Oe applied along the *c*-axis. (b) Magnetic susceptibility data for a second $Co_{0.83}Fe_{0.17}Sn$ crystal with the field applied along both the *c* and the *a* axis. The larger susceptibility for H//*c* may indicate that the Fe spins tend to lie in the plane, as is the case for pure FeSn. (c) Magnetization data for $Co_{1-x}Fe_xSn$ crystals with x = 0.02, 0.08, 0.17 with H//c at temperatures below the temperature, $T_g$. The magnetization data for H//a is the same within experimental error. Also shown (dashed line) is the expected magnetization curve at 0.4 K for non-interacting spins with S = 1 and g = 2.

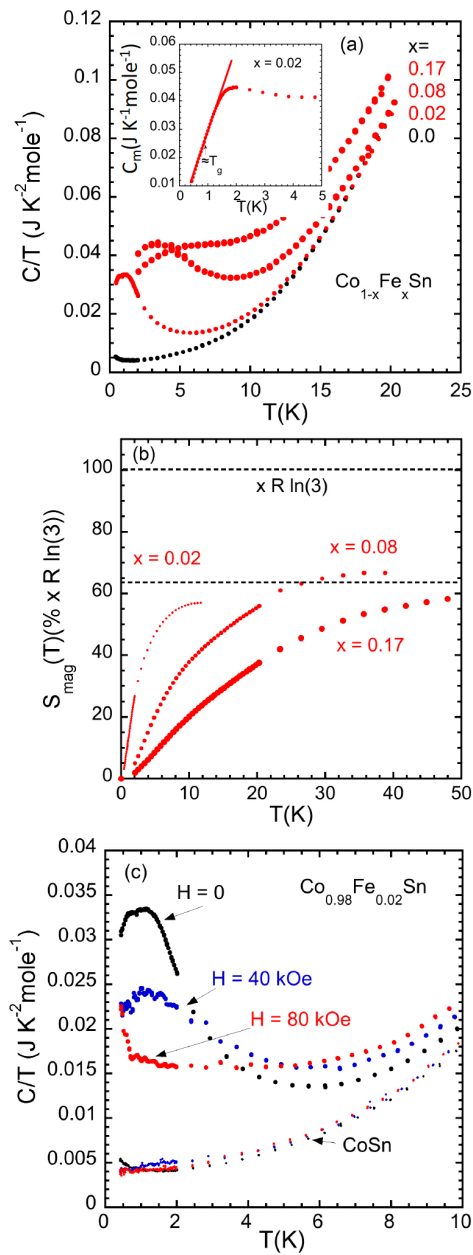

Fig. 6 (a) Specific heat divided by temperature versus temperature for CoSn and 3 iron-doped alloys. Inset: Magnetic specific heat versus temperature for x=0.02 illustrating the linear temperature dependence near and below $T_g \approx 1K$. (b) Magnetic entropy removed due to magnetic order versus temperature. For a concentration x of Fe spins with S=1 the expected entropy is xRln(3). The measured entropy release is about 60% of this value, which is close to the value of xRln(2), as noted by the second dashed line. (c) Specific heat divided by temperature versus temperature for a CoSn crystal and a CoSn crystal doped with x = 0.02, in applied magnetic fields of H= 0, 40, and 80 kOe.

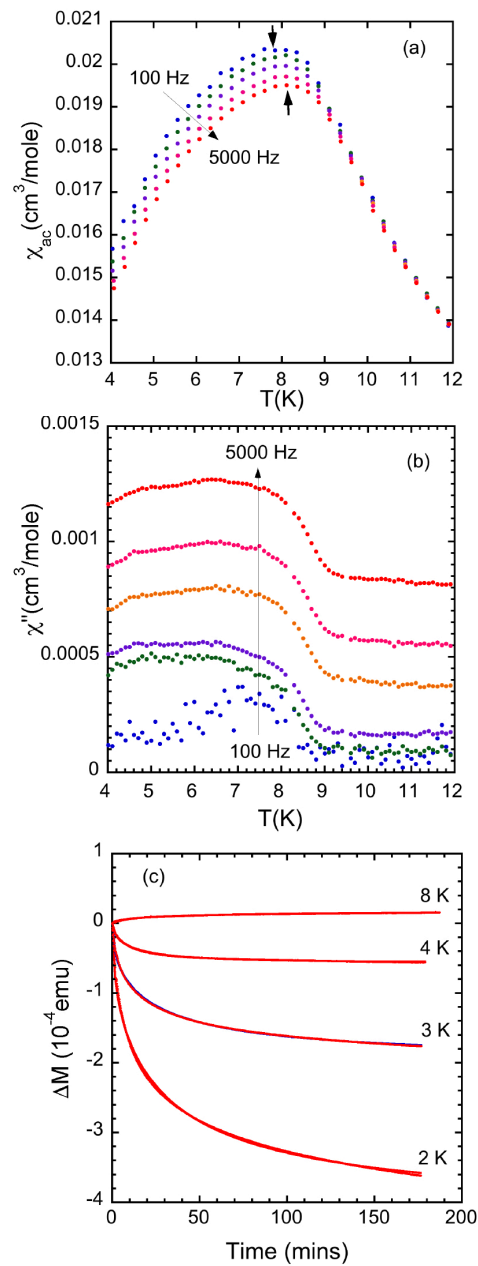

Fig. 7. (a) real part of ac susceptibility versus temperature and frequency for a $Co_{0.83}Fe_{0.17}Sn$ crystal. The shift of the maximum with frequency (black arrows) is indicative of glassy behavior. (b) imaginary part of the susceptibility near Tg. Note the magnitude is much smaller than the real part. (c) change in magnetization versus time for temperatures below glass transition temperature. In this time interval the relaxation can be fit to a stretched exponential. The crystal was cooled from 20 K in a field of 10 kOe to the desired temperature and then the field was set to 0 (which took about 3.5 minutes). The magnetization continued to drift downwards implying some glassy response. For temperatures above $T_g$, the magnetization was essentially independent of time.

*In-doped CoSn crystals*

No magnetic order or superconductivity was observed for the In-doped CoSn crystals. A comparison of the magnetic susceptibility data from three In-doped CoSn crystals ($CoSn_{0.95}In_{0.05}$, $CoSn_{0.9}In_{0.1}$, $CoSn_{0.8}In_{0.2}$), pure CoSn, and a lightly Fe doped crystal ($Co_{0.99}Fe_{0.01}Sn$) is shown in Fig. 8a. The low temperature increase in the susceptibility for the $CoSn_{0.95}In_{0.05}$ crystal is comparable to CoSn but much smaller than for the 1.3% Fe sample. Notice however that the susceptibility of the In doped samples at room temperature are about three to 3-12 times larger than pure CoSn. At 2 K the magnetization curves, M(H), for the three In-doped crystals are linear (not shown).

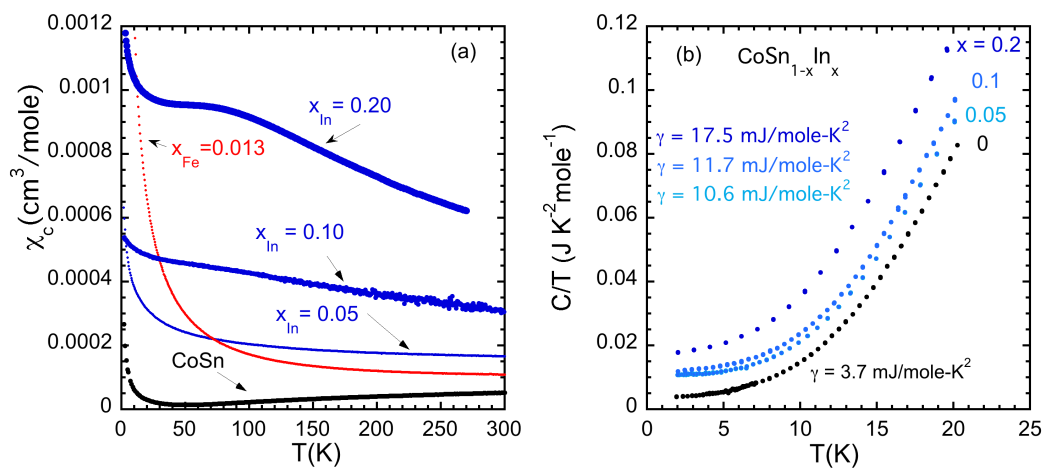

Fig. 8 (a) Comparison of the magnetic susceptibility data from CoSn, $Co_{0.987}Fe_{0.013}Sn$, $CoSn_{0.95}In_{0.05}$, $CoSn_{0.90}In_{0.1}$. and $CoSn_{0.8}In_{0.2}$ with H = 10kOe. Notice that the room temperature susceptibility of all of the In-doped samples are larger than the value for pure CoSn. This suggests a larger Pauli susceptibility with indium doping and presumably a corresponding larger density of states. (b) C/T versus T for three In-doped crystals and CoSn.

Specific heat data from the In-doped CoSn crystals are shown in Fig. 8b. Although the In-doped crystals did not show any evidence of magnetic order or superconductivity, the value of the Sommerfeld coefficients were about 3-5 times larger than the value for pure CoSn. The larger $\gamma$ is consistent with the larger Pauli susceptibility inferred from Fig 8a. An applied magnetic field of 80 kOe had no measurable effect on the low temperature specific heat of the In-doped crystals.

*Ni-doped CoSn crystals*

The addition of Fe or In to CoSn should add holes and move the flat bands closer to the Fermi energy. By contrast, Ni-doping of CoSn should add electrons and move the flat bands further away from $E_F$. A comparison of the magnetic susceptibility and low temperature specific heat data from $Co_{0.93}Ni_{0.07}Sn$ and CoSn are displayed in Fig. 9. Both the susceptibility and the Sommerfeld coefficient are smaller for the Ni-doped sample consistent with this hypothesis. Even though the center of the flat bands in CoSn are about 100 meV below the Fermi energy, spin orbit coupling and interactions beyond nearest neighbor broaden the bands. The DFT calculations in Fig 1d, indicate that the d-bands nearest $E_F$ have a width of about 0.3 eV, with part of the spectral weight from the d-bands extending up to $E_F$, even for pure CoSn.

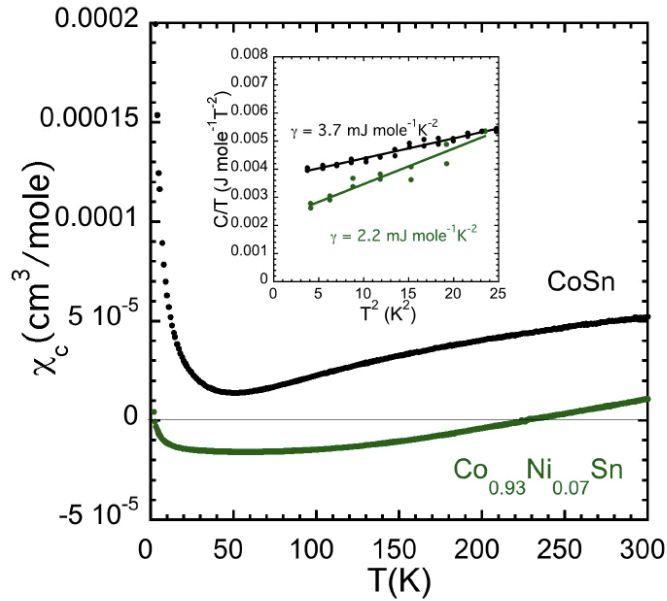

Fig. 9 Comparison of the magnetic susceptibility of pure CoSn and a Ni-doped CoSn crystal, $Co_{0.93}Ni_{0.07}Sn$ taken with H = 10 kOe. Low temperature specific heat data are shown in the inset for the same two crystals.

*Resistivity*

The effect of Fe, In and Ni doping on the c-axis resistivity, $\rho_c$, of CoSn is shown in Fig. 10. All of the crystals are extremely good metals and the disorder caused by doping increases the residual resistivity of all of the alloys relative to pure CoSn . Replacing 2% of the Co with Fe results in significantly more scattering than replacing 10% of the Sn by In. For higher concentrations of Fe (x=0.08, 0.17), $\rho_c$ exhibits a positive curvature, similar to that found in pure FeSn [7,8]. There is no anomaly in the resistivity in the vicinity of $T_g$. Although not as noticeable, all of the resistivity curves from the In-doped crystals also exhibit a positive curvature for temperatures above 100 K. A positive curvature in resistivity curves will occur if there is a sharp increase in the density of states near the Fermi energy [21]. By contrast, the resistivity data from the Ni doped crystal ($Co_{0.93}Ni_{0.07}Sn$) exhibits negative curvature over the 2-300 K temperature range, suggesting that the higher density of states associated with the flat bands are further from $E_F$. Also, doping with Ni *reduces* the room temperature resistivity relative to pure CoSn, although the residual resistivity is increased from about 0.5 $\mu\Omega$-cm to 2 $\mu\Omega$-cm for CoSn and $Co_{0.93}Ni_{0.07}Sn$, respectively.

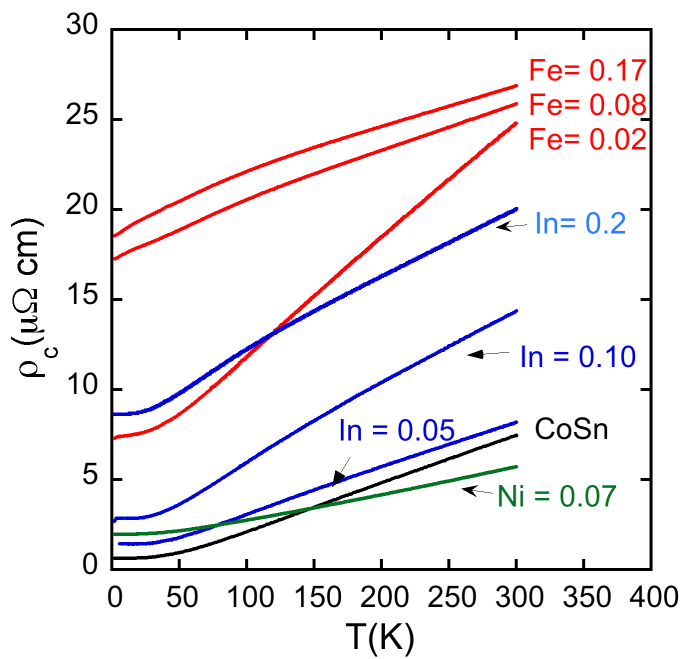

Fig. 10 Resistivity vs temperature for CoSn crystals doped with Fe, In or Ni.

**Discussion**

The flat d bands in pure CoSn are about 100 meV below the Fermi energy as determined from angle-resolved photoelectron spectroscopy [11,12], magnetic susceptibility and theory [13]. For low doping concentrations of order x = 0.01, a rigid band picture often works [22] and each Fe or In atom should create 1 hole. Assuming a single parabolic band and effective masses of 2-5 $m_0$ that are typical of d bands in these materials [7,10], the estimated shift of the Fermi energy at a doping level of x = 0.01 is between 30 and 80 meV. This is a rough estimate since there are also lighter bands (e.g. with smaller effective mass) near $E_F$, but for the lighter bands a similar doping level will move the Fermi energy much further, by 160 meV or more. At higher doping levels the rigid band model begins to break down and in general, doping most materials is quite complex due to lattice strain effects, the formation of compensating defects, trapping of charge, etc. However, these rough calculations suggest that doping CoSn with a few percent of Fe or In should move the Fermi energy into the flat bands (which from DFT have a width of about 0.3 eV, see Fig. 1d). In principle, Hall measurements could be used to estimate the change in carrier concentration with doping, but since the alloys are excellent metals (Fig. 10) with multiple bands near the Fermi energy [13, Fig. 1d], trying to interpret small signals from a Hall measurement is not straightforward.

From this simple picture it should not matter where the holes come from, Fe or In. We expect the density of states at $E_F$ should increase at a similar rate. This appears to be true from our heat capacity data shown in Figs. 6 and 8. Doping with In significantly increases the density of states (value of $\gamma$) at $E_F$ by about a factor of 3-5 relative to pure CoSn (Fig. 8b). For the Fe-doped crystals, there is likely a similar increase in the density of states although it is difficult to extract a $\gamma$ value cleanly from the specific heat data shown in Fig. 6 because of the magnetic transition.

The large increase in the density of states with In doping is apparently not enough to produce magnetic order via a Stoner-like mechanism [23]. The magnetic susceptibility of the $CoSn_{0.8}In_{0.2}$ crystal does show a broad plateau near 75 K (Fig. 8a) but there is no obvious phase transition associated with this feature. There is no separation between zfc and fc susceptibility data taken with a small applied field of 100 Oe (not shown). Also as noted previously, magnetization versus field measurements at 2 K are linear for all of the In-doped samples including $CoSn_{0.8}In_{0.2}$ unlike the results obtained for the Fe-doped crystals (Fig. 5c).

By contrast the Fe-doped crystals, $Co_{1-x}Fe_xSn$, exhibit some type of magnetic transition, which we believe to be a spin glass, for values of x as small as 0.02. Doping with Fe not only adds holes to CoSn, but each Fe atom also has a local magnetic moment of about 2 $\mu_B$, which we have used to accurately determine the Fe concentration in our crystals (Fig 3). These moments likely couple with one another via some type of RKKY-like interaction resulting in the formation of a spin glass [24]. The local magnetic moment associated with Fe also has a significant effect on electrical transport, adding an additional scattering mechanism that results in much higher residual resistivities (relative to In-doping) and a positive curvature in $\rho_c(T)$ at higher temperatures (Fig. 10).

Several key bulk measurements are associated with "canonical" spin glasses [24] and the $Co_{1-x}Fe_xSn$ exhibit all of these features at least for $0.02 < x < 0.2$ :
- shift of peak in ac susceptibility data with measuring frequency (Fig. 7a)
- separation of zfc and fc dc susceptibility data near the spin-glass temperature, $T_g$ (Fig. 5a)
- time dependence of dc magnetization data below $T_g$ (Fig. 7c)
- a broadened peak in specific heat near $T_g$ with a substantial portion of the spin entropy recovered well above $T_g$ (Fig. 6)
- a magnetic specific heat that is linear in T below $T_g$ (inset Fig. 6a).

The ac susceptibility data of $Co_{0.83}Fe_{0.17}Sn$ shows a frequency dependence (Fig. 7a) and the peak, $T_p$, in the real part of the ac susceptibility shifts by about 0.25 K for a change in frequency from 100 hz to 5000 hz. The small shift in $\Delta T_p/(T_p \Delta \log_{10}(\omega)) \approx 0.007$ is in the range of canonical spin glasses that have strongly interacting cooperative freezing [24]. However, the peak in the ac susceptibility data is broader than in classic spin glasses such as Cu:Mn [26]. Also for many spin glasses the width of the ac susceptibility peak is rapidly broadened in small applied DC magnetic fields [25,26]. This is not the case for the $Co_{0.83}Fe_{0.17}Sn$ sample. The data shown in Fig. 7a were taken with $H_{DC} = 0$ or H =100 Oe and there was no significant change in the width of the peak. Part of this may be a result of the variation of the local Fe concentration in each crystal (see crystal growth and chemical characterization). We note, however, that even for the case of Cu:Mn, the sharpness of the ac susceptibility feature depended on details of the synthesis conditions [26].

The separation between the zfc and fc susceptibility data taken with temperature sweep rates of 1 K/min are shown for several compositions in Fig 5a. The separation temperatures were used to construct the phase diagram in Fig. 2 for several values of x < 0.6. The time dependence of the magnetization below $T_g \approx$ 6K for a $Co_{0.83}Fe_{0.17}Sn$ crystal is shown in Fig. 7c. For temperatures below $T_g \approx$ 6 K, the magnetization continues to decrease for several hours. For the time interval shown in Fig. 7c the data can be fit or parameterized by a stretched exponential function of the form $A(1-exp(-t/\tau)^\beta)$. The stretched exponential function occurs often in parametrization of the glass transition of structural glasses and can be thought of as a distribution of relaxation times with $0 < \beta < 1$ [27,28]. The smaller the value of $\beta$, the broader the distribution of relaxation times. The isothermal changes in magnetization versus time in the time interval shown are accurately described by a stretched exponential decay function with $\tau$ = 30.3, 25.0 ,12.2 mins and $\beta$ = 0.51, 0.51, 0.57 for T = 2, 3 and 4 K respectively. This also indicates that the M(H) curves shown in Fig. 5c will depend somewhat on the rate at which the data are taken, and previous thermal history.

The broad peak in the magnetic contribution to the specific heat data near $T_g$, and the general behavior of the peak in a magnetic field (see Fig 6) are similar to heat capacity data reported for canonical spin glasses such as Cu:Mn [24]. Taken together, all of the magnetic and specific heat measurements imply that the feature labeled as $T_g$ in the phase diagram shown in Fig. 2 denotes a transition to a spin glass phase. This interpretation is also consistent with both single crystal and powder neutron diffraction results on $Co_{0.83}Fe_{0.17}Sn$ that detected no indication of long-range antiferromagnetic order at 1.6 K. The phase diagram of $Co_{1-x}Fe_xSn$ in Fig.2, which combines our previous work on the Fe rich alloys [9] with the present work on the lightly doped Fe alloys, also indicates that a magnetic quantum critical point for values of x near 0.4 is avoided by the appearance of this spin glass phase.

The substitution of Ni into CoSn should add electrons and move the flat bands further away from $E_F$. As shown in Fig. 9, the decrease in both the magnetic susceptibility and $\gamma$ relative to the values for CoSn are consistent qualitatively with this idea. To get an approximate idea of the position of the *edge* of the flat d bands in $Co_{0.93}Ni_{007}Sn$ with respect to $E_F$ we analyze the magnetic susceptibility data in Fig. 9 using a simple model that we previously used for pure CoSn [13]. That model accounts for the contribution of the flat band to the Pauli susceptibility. The toy model treats the effect of the flat band as a step function in the density of states at a position $E_{step}$ below $E_F$. As the temperature is increased from T = 0, the Pauli susceptibility will begin to increase when the temperature is high enough to have electrons contribute from the flat band. From this analysis the edge of the flat band, $E_s$, for CoSn is about 20 meV while for the $Co_{0.93}Ni_{0.07}Sn$ crystal $E_s \approx$ 80 meV. For more details see [13]. The effect of moving the flat bands further away from $E_F$ with Ni-doping is also reflected in the resistivity. The alloying of Ni into CoSn should increase disorder scattering and it does increase the residual resistivity from 0.5 $\mu\Omega$-cm for CoSn to 2 $\mu\Omega$-cm for $Co_{0.93}Ni_{0.07}Sn$. Remarkably, however, at room temperature the resistivity of the Ni-doped crystal is 5 $\mu\Omega$-cm, less than for pure CoSn, which has a value near 7 $\mu\Omega$-cm. This suggests that the flat bands associated with Co d orbitals scatter more as they are tuned to $E_F$.

## Conclusions

The effects of electron or hole doping on the magnetic, thermodynamic and transport properties of CoSn are reported. Hole doping with a few percent of Fe or In should move the flat bands closer to $E_F$, whereas electron doping with Ni should move the flat bands further away from $E_F$. We provide evidence that this is indeed what occurs, but the resulting ground state is different for Fe versus In doping. Hole doping with Fe increases the density of states at $E_F$, and results in a spin glass for at least $0.02 < x < 0.2$, in $Co_{1-x}Fe_xSn$ single crystals. Hole doping with indium results in large increases in the Sommerfeld coefficient $\gamma$ and the Pauli susceptibility, but no clear evidence of magnetic order or a spin glass is observed. Crystals with compositions near the In solubility limit, $CoIn_{0.2}Sn_{0.8}$, exhibit a broad plateau in the magnetic susceptibility data near 75 K and a large Sommerfeld coefficient of 17.5 $mJ/K^2$-mole. Future work will use neutron scattering to more carefully examine these In-doped crystals for possible magnetic order. Electron doping with Ni results in a smaller Pauli susceptibility and Sommerfeld coefficient, consistent with the expectation that the flat d bands are further away from the Fermi energy than in pure CoSn. A simple model that accounts for the increase of the Pauli susceptibility with increasing temperature gives the position of the of the flat bands below $E_F$ as 20 meV for CoSn and 80 meV for $Co_{0.93}Ni_{0.07}Sn$.

## Acknowledgements

This research was supported almost entirely by the U. S. Department of Energy, Office of Science, Basic Energy Sciences, Materials Sciences and Engineering Division. Help with the neutron scattering experiments and data interpretation was provided by YHL, MBS, and QZ of the Neutron Scattering Division. This research used resources at the Spallation Neutron Source, a DOE Office of Science User Facility operated by Oak Ridge National Laboratory.